\documentclass[12pt]{article}

\usepackage{amsmath}    
\usepackage{amssymb}
\usepackage{graphicx}
\usepackage{amsfonts}  
\usepackage{epsfig}
\usepackage{graphics}
\usepackage{amssymb,amsmath}
\usepackage{amsthm}
\usepackage{amsfonts}  

\title{\bfseries\huge  \vspace{-3.0cm} On the constancy of light speed}
\author{ \itshape by \\ \\ \bfseries Abdul Rouf \\\\ National Institute of Technology, Srinagar, Kashmir. \\\normalsize Email: abdulrouf.lightspeed@yahoo.com}

\begin{document}
\date{}

  \maketitle 
  
  \begin{abstract}
	Method given for intuitively visualizing constancy of light speed.
Using mass-energy equivalence, we can show with the help of simple thought
experiments, that velocity time dilation and gravitational time dilation are a
necessary consequence of principle of relativity. Speed of light is discussed in
the context of each kind of time dilation.

 \end{abstract}

  \section{Introduction}
  
  \hspace{.5cm} We know light is an electromagnetic wave. So according to Maxwell's equations the speed of light in vacuum [1] $c_0 = 1/\sqrt{\mu_0\epsilon_0}$  (=299 792 458 m/s) [2][3]. In any given inertial frame, the speed of light is a constant, independent of the state of motion of the emitting body. This fact has been confirmed by experiments [4][5][6][7]. Albert Einstein in his paper [8] `On the Electrodynamics of Moving Bodies', which was published in 1905 and gave birth to special relativity, used constancy of light speed as a postulate.\\

Even though theoretically proved and experimentally verified, students of physics find the constancy of light speed counterintuitive and logically inconsistent. The main reason is that, in introductory courses to special relativity, no or insignificant explanation is given for its intuitive understanding. No physical mechanism is provided which can satisfactorily account for constancy of light speed. Instead, it is directly introduced as a fundamental postulate in special relativity. Consequently, other relations derived using constancy of light speed appear `mysterious'. Hence, it will be highly advantageous to have a physical interpretation of constancy of light speed and comprehend its underlying physics, before we proceed to show its implications in special relativity. This \vspace{-1.0cm}paper is an attempt in this direction. \\

I will start with stating principle of relativity and deriving the mass energy equivalence equation $ E=mc^2 $ by Einstein's Gedankenexperiment [9] [10]. I will use them in a thought experiment to deduce velocity time dilation, which can subsequently account for constancy of light speed. Then using similar method in another thought experiment, gravitational time dilation will be deduced and its effect on speed of light will be discussed. Finally, the paper is concluded. Throughout this paper, only classical physics is used. This paper, it is anticipated, will be useful for understanding and also for teaching the foundations of special relativity.\\

  \section{Starting points}
  
  \subsection{\normalsize The Principle of Relativity } The laws of physics are the same for all observers in uniform motion relative to one another. Experiments will have same results in every inertial frame and thus can't be used to determine absolute velocity of any inertial frame.  \\
  
  \subsection{\normalsize Einstein's Gedankenexperiment } We suppose [11] that an amount $E$ of radiant energy (a burst of photons) is emitted from one end of a box of mass $M$ and length $L$ that is isolated from its surroundings and is initially stationary. According to classical physics, the radiation carries momentum $E/c$. Since the total momentum of the system remains equal to zero, the box must acquire a momentum equal to  $-E/c$.\\ 
\begin{figure}[htbp]
	\centering 
		\includegraphics[width=0.95\textwidth]{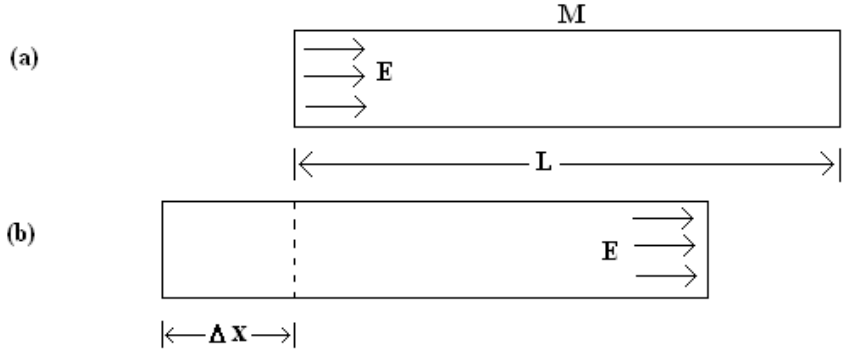}
		\vspace{-2.0cm}\caption{\itshape Einstein's box - a hypothetical experiment in which a box recoils from its initial position (a) to a final position (b) as a result of a burst of radiant energy from one end of the box to the other.}

\end{figure}

 Hence the box recoils with a speed $v$, given by {\large \begin{equation}
 v = - \frac{E}{Mc} \end{equation}}\\

\vspace{-0.2cm}After traveling freely for a time  $ \Delta t $ ( = $L/c$ very nearly, provided $ v << c $), the radiation hits the other end of the box and conveys an impulse, equal and opposite to the one it gave initially, which brings the box to rest again. 

 \hspace{-0.2cm}Thus the result of this process is to move the box through a distance $\Delta x$ :

{\large \begin{equation}\Delta x =v\Delta t =  - \frac{EL}{Mc^2}
 \end{equation}}\\

\vspace{-1.0cm}But this being an isolated system, we are reluctant to believe that the center of mass of the box plus its contents has moved. We therefore postulate that the radiation has carried with it the equivalent of a mass $m$, such that {\large\begin{equation}
mL + M\Delta x = 0
 \end{equation}}\\\\
 
 \vspace{-1.5cm}Putting Eq.(2) and Eq.(3) together, we have {\large \begin{equation}
 m=\frac{E}{c^2} \hspace{.5cm} or  \hspace{.5cm}  E=mc^2
 \end{equation}}\\
  
  \vspace{-.8cm}This is the famous mass-energy equivalence equation, which was proposed by  Einstein in 1905, in the paper [12] `Does the inertia of a body depend upon its energy-content?'. It states that energy equals mass multiplied by the speed of light squared.

   \section{Velocity time dilation }
  
  \hspace{.5cm} Let $S$ and  $ S^\prime$ be the two inertial frames of reference. Let there be two observers Mr.X and Mr.Y in the frames $S$ and $ S^\prime $ respectively. Also, let there be a Hydrogen atom in the frame $ S^\prime $. Consider that initially both the frames, $S$ and  
  $ S^\prime $, were at rest with respect to each other.\\

  Imagine that high frequency radiations are incident on both the frames along X-axis. Let its frequency be $ \nu_0 $. Let the energy  $ h\nu_0 $  ( where $h$ is Plank's constant ) of the radiations, as measured by the observers, be just equal to the ionization energy of the hydrogen atom. So for both the observers ionization of the H-atom will occur.\\ \\
  
   \vspace{-1cm}\begin{figure}[htbp]
	\centering
		\includegraphics[width=1.04\textwidth]{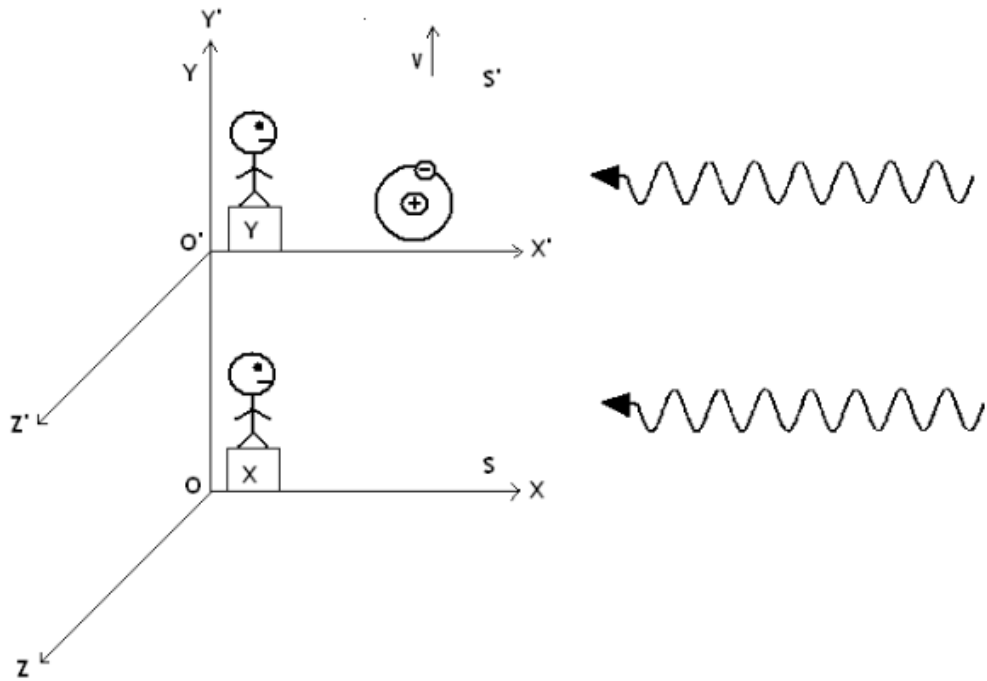}
	\caption{\itshape Mr.X and Mr.Y are in frames $S$ and $ S^\prime $ respectively. Frame $S^\prime$ moves with respect to frame $S$ with a uniform velocity $v$ in the positive direction of Y-axis. High frequency radiations are incident on both frames along X-axis. }
	
\end{figure}

 Now let us suppose that  frame $ S^\prime $ moves with respect to frame $S$, with a uniform velocity $v$, in the positive direction of Y-axis ( i.e. ${\perp}$ to the direction of propagation of the incident radiations ). We can show [13] that $ E = mc^2 $ leads to the growth of inertia with velocity and consequent increase in mass as  $ m = m_0 \gamma $ (where $ \gamma $ is the Lorentz factor). Thus, if $ m_0 $ was the rest mass of the electron, then its relativistic mass $m$ [14][15] as apparent to Mr.X  will be : 
   {\large\begin{equation}
   m = \frac{m_0}{\sqrt{1-\frac{v^2}{c^2}}}
 \end{equation}}

If the rest mass of electron is $m_0$, the binding energy $E_0$ of the ground sate of the Hydrogen atom is : 

{\large\begin{equation}
E_0 = - \frac{m_0k^2}{2\hbar^2} \hspace{1cm} \left[ where \hspace{.2cm}  k = \frac{e^2}{4\pi \epsilon_0} \right] \end{equation}}\\

\vspace{-.8cm}Taking the increase in mass of the electron as shown in Eq.(5) into account, the  binding energy $E$ of the ground state of the Hydrogen atom in frame $S^\prime$ as apparent to Mr.X  will be : {\large\begin{equation}
 E = - \frac{m_0k^2}{2\hbar^2\sqrt{1-\frac{v^2}{c^2}}}\Rightarrow   E=\frac{E_0}{\sqrt{1-\frac{v^2}{c^2}}}\end{equation}}\\ 
 
 \vspace{-.8cm}This means that the binding energy of  H-atom will increase by Lorentz factor due to increase in mass of its electron. Because ionization energy is equal to binding energy in magnitude, therefore for Mr.X, ionization energy of H-atom will also increase by Lorentz factor. However, frequency and thus energy of the radiations will remain same for Mr.X. Therefore for Mr.X, now energy of the radiations will be less than the new ionization energy of  H-atom. Thus for Mr.X, H-atom in  frame $S^\prime$  will not ionize.\\\\\\
 
 \vspace{-.8cm}Now let us see what Mr.Y will observe. Because Mr.Y is at rest with respect to H-atom, therefore he will not observe any increase in the mass of the electron. Thus for him ionization energy of H-atom will not increase.\\
 
  Now if we assume that for Mr.Y, frequency and thus energy of the radiations perpendicular to his motion will remain same, then this would mean that for Mr.Y, H-atom will ionize. But this can't be true, because Mr.X and Mr.Y can't observe different results for the same experiment, for that violates principle of relativity.\\

The only way both observers can observe same results is that for Mr.Y, frequency of the incident radiation must decrease by the same factor by which ionization energy of H-atom increased for Mr.X  i.e. Lorentz factor. So for Mr.Y, apparent frequency $ \nu $ will be :   \vspace{-.2cm} {\large \begin{equation}
\nu = \nu_0\sqrt{1-\frac{v^2}{c^2}} \end{equation}}\\

\vspace{-.8cm}We know $\nu = 1/t$ (where  $t$ is the time period). Thus, if $t_0$ be the proper time period, then from Eq.(8) we have apparent time period $t$ as : \\ {\large \begin{equation}
 t = \frac{t_0}{\sqrt{1-\frac{v^2}{c^2}}}\end{equation}}\\
 
 \vspace{-.8cm}This expresses the fact that for the moving observer the period of the clock is longer than in the frame of the clock itself. It is called velocity time dilation. This effect is reciprocal i.e as observed from the point of view of any two clocks which are in motion with respect to each other, it will be the other party's clock that is time dilated.

   \section{Velocity time dilation and light speed}
 
  \hspace{.5cm} Constancy of light speed means that the speed of light in free space has the same value in all inertial frames of reference. In this section, I intend to prove that constancy of light speed can be shown as a direct consequence of velocity time dilation.\\\\
 
  \vspace{-.5cm} As concluded in section $3$ above, a clock that moves with respect to an observer ticks more slowly than it does without such motion. Now let us consider a hypothetical clock called photon clock. In it light is reflected back and forth between its mirrors and whenever the light strikes a given mirror, the clock ticks once. If this clock is in inertial motion with respect to an observer, then velocity time dilation will cause it, like all other kinds of clocks, to tick slower. However, because the clock is moving, the light pulse will trace out a longer, angled path between the mirrors. The net result of velocity time dilation and increase in the path length is that the speed of light in the moving photon clock remains equal to the speed of light in the rest photon clock. In other words, speed of light remains constant.\\

To prove above explanation, let $c$ be the speed of light in a photon clock which is at rest relative to an observer on ground and $ c_0 $ be the speed of light in an identical photon clock moving at the speed $v$ relative to the observer on ground. Clearly, both the clocks are in inertial frames of reference.\\

\vspace{-.6cm}\begin{figure}[htbp]
	\centering
		\includegraphics[width=1.139\textwidth]{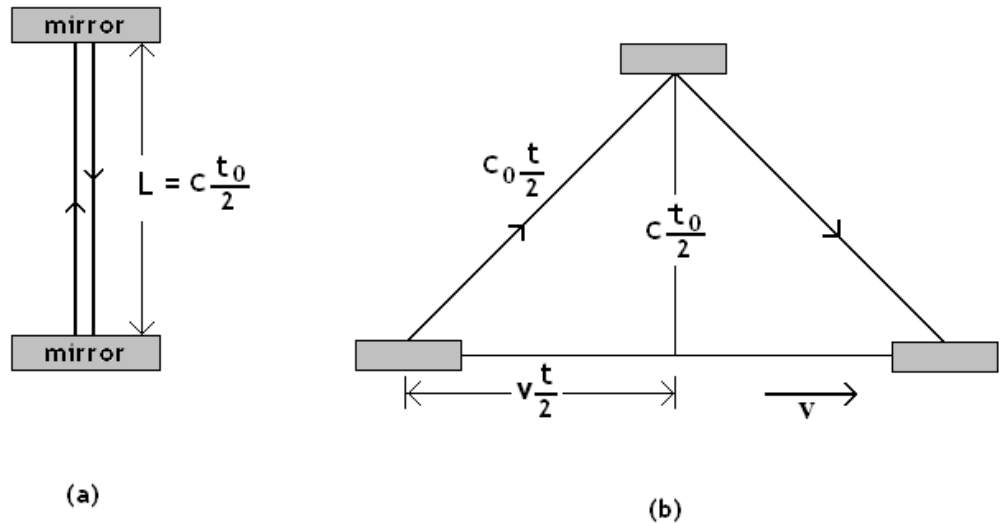}
		\vspace{-2cm}\caption{\itshape(a) A photon clock at rest relative to an observer on ground.
		(b) An identical photon clock moving at the speed v relative to the observer on ground.}

\end{figure}

\vspace{-.2cm}In the photon clock at rest , the time interval between ticks is the proper time $ t_0 $ and the time needed to travel between the mirrors at the speed of light $c$ is $ t_0/2 $. Hence the vertical distance $L$ between its mirrors (and the mirrors of identical moving photon clock) is equal to $ c(t_0/2) $.  \\

\vspace{.3cm}In the moving photon clock, the mirrors are parallel to the direction of its motion. Let $t$ be the dilated time interval between its ticks. Because the clock is moving, the light pulse, as seen by the observer on the ground, follows a zigzag path. On its way from the lower mirror to the upper one in the time $t/2$, the light pulse travels a horizontal distance $v(t/2)$ and a total distance $ c_0(t/2) $. Since $ c(t_0/2) $ is the vertical distance between its mirrors, using Pythagorean theorem we have :

{ \large \begin{equation}
	\hspace{-2cm}\vspace{-.2cm} \left( \frac{ct_0}{2}\right)^2 = 	\left( \frac{c_0t}{2}\right)^2 - 	\left(  \frac{vt}{2}\right)^2 \end{equation}

\begin{equation}
		\hspace{-2cm} \vspace{-.2cm} c^2\left( \frac{t_0}{2}\right)^2 = 	\left( \frac{t}{2}\right)^2 \left(c_0^2 - v^2\right)\end{equation}

\begin{equation}
			\hspace{-2cm} \vspace{-.2cm}c^2\left( \frac{t_0}{t}\right)^2 =  c_0^2 - v^2\end{equation}

\begin{equation}
	  \hspace{3.5cm} \vspace{-.3cm} c^2\left(1-\frac{v^2}{c^2}\right)	=  c_0^2 - v^2  \hspace{.8cm}\left[ \hspace{.05cm}using\hspace{.05cm}Eq.(9) \hspace{.05cm} \right]\end{equation}

\begin{equation}
 	\hspace{-2cm}	\vspace{-.6cm} c^2 - v^2 = c_0^2 - v^2 \end{equation}

	\begin{equation}
 	\hspace{-2cm}\vspace{-.5cm}	c = c_0 \end{equation}}


Thus, speed of light in rest photon clock  is equal to the speed of light in moving photon clock or in other words speed of light is constant. This proves that velocity time dilation can account for constancy of light speed .\\

The case discussed above, in which the mirrors of photon clock are parallel to the direction of its motion, is a simple one. Here velocity time dilation alone can account for constancy of light speed. Now if the photon clock is rotated, such that its mirrors are no more parallel to the direction its motion, then the distance between the mirrors $L$ will undergo length contraction [16]. In such cases, in order to account for light speed constancy, in addition to velocity time dilation, we will have to include the effect of length contraction as well. However, length contraction itself can again be deduced from velocity time dilation and can be easily shown as a necessary consequence of velocity time dilation and the principle of relativity [17]. Thus velocity time dilation can still be considered the fundamental cause of light speed constancy.

  \section{Gravitational time dilation}

\hspace{.5cm}Suppose there are two observers Mr.X and Mr.Y on the surface of a massive body (i.e. lower gravitational potential). Let there be a hydrogen atom with them. Consider that they make radiations of frequency $ \nu_0 $ incident on it with the help of a radiation source on the ground. Let its energy,  as measured by both the observers, be just equal to the ionization energy of the H-atom. So for both the observers ionization of the H-atom will occur.\\

Let us consider the position of an observer as his `zero of gravitational potential energy'. Now suppose  Mr.Y goes up to a higher altitude $h$ along with H-atom (to higher gravitational potential). Now for Mr.X, electron of the H-atom will gain potential energy $ P.E =  m_0gh $ (where $ m_0 $ is electron mass, $g$ is acceleration of gravity). Using the result of Gedankenexperiment $ E=mc^2 $, the new mass $m$ of the H-atom's electron as apparent to Mr.X will be :

 {\large\begin{equation}
m = m_0 + m_0\frac{gh}{c^2} = m_0 \left(1+\frac{gh}{c^2}\right)
 \end{equation}}
 
 Taking the increase in mass of the electron as shown in Eq.(16) into account, binding energy $E$ of the ground state of the Hydrogen atom in higher gravitational potential as apparent to Mr.X will be :
 
 {\large \begin{equation}
E =  - \frac{m_0 \left(1+\frac{gh}{c^2}\right)k^2}{2\hbar^2} \Rightarrow  
E = E_0\left(1+\frac{gh}{c^2}\right) \end{equation}}\\

This means that the binding energy of H-atom will increase by $ 1 + gh /c^2 $  due to increase in mass of its electron. Consequently this would mean that for Mr.X, ionization energy of H-atom will increase by $ 1+ gh /c^2. $ However, frequency and thus energy of the radiation will remain same for Mr.X. Therefore for Mr.X, now energy of the radiation will be less than the new ionization energy of H-atom. Thus for Mr.X, H-atom at the high altitude will not ionize with the same radiation source. \\

\begin{figure}[htbp]
	\centering
		\includegraphics[width=1\textwidth]{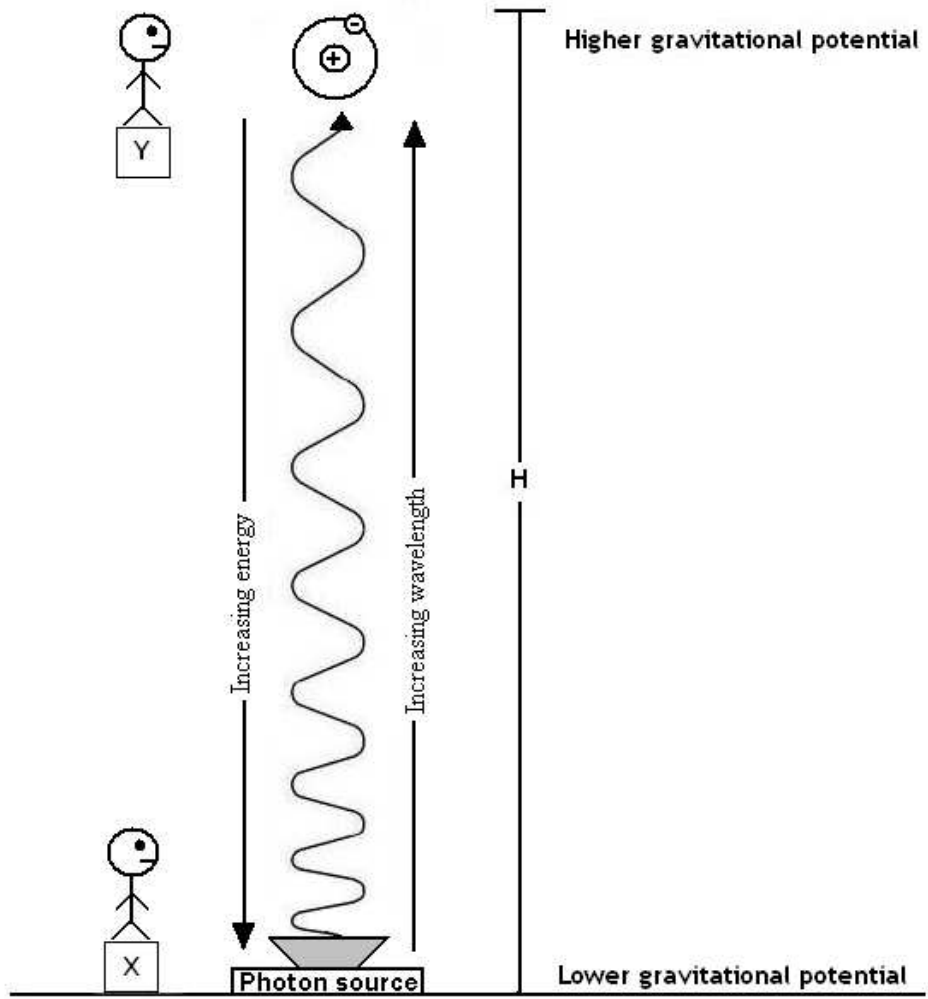}
		\caption{\itshape Mr.X is in the lower gravitational potential and Mr.Y, along with
 H-atom, in the higher gravitational potential. Mr.Y observes a decrease in frequency of radiation coming from Mr.X.}

\end{figure}

\vspace{-.8cm}Now let us see what will Mr.Y observe. Because Mr.Y is at the same altitude as H-atom, therefore for him , its electron will not gain any potential energy. Consequently he will not observe any increase in the mass of the \vspace{-.2cm} electron. Thus for him ionization energy of H-atom will not increase.\\

 Now if we assume that for Mr.Y, frequency and thus energy of the radiation coming from Mr.X (using same radiation source) will remain same, then this would mean that for Mr.Y, H-atom will ionize. But this contradicts principle of relativity.\\

The only way both observers can observe same results is that for Mr.Y, frequency of the incident radiation must decrease by the same factor by which ionization energy of H-atom increased for Mr.X i.e. $1 + gh/c^2$. So for Mr.Y, apparent frequency $\nu$ will be : 

\vspace{-.3cm}{\large\begin{equation}
\nu = \frac{\nu_0}{\left(1+\frac{gh}{c^2}\right)} \end{equation}}

If $t_0$ be the local proper time period, then from Eq.(18) we have apparent time period $t$ as :

{\large\begin{equation}
 t = t_0 \left(1+\frac{gh}{c^2}\right)	\end{equation}
}

This shows that clocks which are far from massive bodies (or at higher gravitational potentials) run faster, and clocks close to massive bodies (or at lower gravitational potentials) run slower. It is called gravitational time dilation. This effect is also manifested in accelerated frames of reference by virtue of the equivalence principle. \\

  \section{Gravitational time dilation and light speed}

\hspace{.5cm}The propagation of light is influenced by gravitation [18]. Although speed of light is constant in inertial frames of reference in special relativity, it can vary based on its position for accelerated frames of reference in special relativity and in general relativity. 

\begin{figure}
	\centering
		\includegraphics[width=0.79\textwidth]{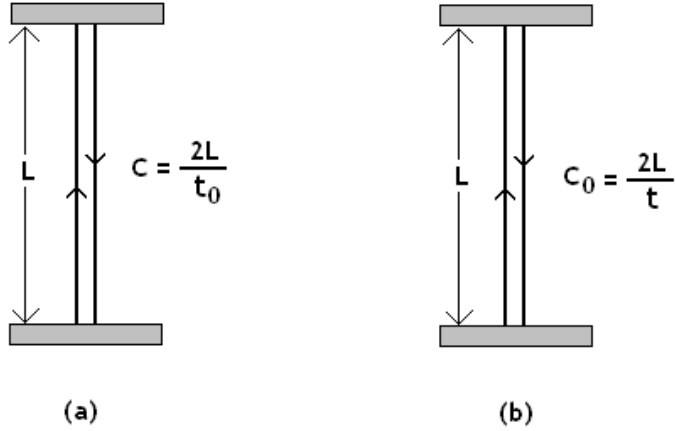}
		\caption{ \itshape (a) A photon clock in higher gravitational potential (b) An identical photon clock in lower gravitational potential.}

\end{figure}

\vspace{3cm}Concerning this fact Einstein wrote [19]
\begin{quote}
	``{\itshape...according to the general theory of relativity, the law of the constancy of the velocity of light in vacuo, which constitutes one of the two fundamental assumptions in the special theory of relativity and to which we have already frequently referred, cannot claim any unlimited validity. A curvature of rays of light can only take place when the velocity of propagation of light varies with position.}"
\end{quote}

To study the effect of gravitational time dilation on the speed of light, let us consider two identical photon clocks (ideally infinitesimal photon clocks so that the curvature of space between its mirrors due to gravitation is negligible) with their mirrors $L$ apart. Let one of the clocks be at a place with gravitational potential $gh$ (i.e higher gravitational potential) and other at a place with zero gravitational potential (i.e lower gravitational potential).Clearly, the time interval between ticks of the photon clocks in higher gravitational potential and lower gravitational potential will be equal to $t_o$ and $t$ [ as used in Eq.(19) ] respectively.\\\\\\

\hspace{-1cm}The speed of light $c$ in the photon clock at gravitational potential $gh$  will be :\\

{\large\begin{equation}
	c=\frac{2L}{t_0}
\end{equation}}\\

\hspace{-1cm} Now the speed of light $c_0$ in the photon clock at zero gravitational potential will be:\\

{ \large\begin{equation}
\hspace{-1.5cm}	c_0 = \frac{2L}{t} \end{equation}

\begin{equation}
\hspace{4.2cm}	c_0 = \frac{2L}{t_0 \left(1+ \frac{gh}{c^2} \right )} \hspace{.5cm}\left[ \hspace{.1cm}using\hspace{.1cm}Eq.(19) \hspace{.1cm} \right]\end{equation}

\begin{equation}
\hspace{4.6cm} c_0 = \frac{c}{ \left (1+\frac{gh}{c^2}\right)} \hspace{.5cm}\left[ \hspace{.1cm}using\hspace{.1cm}Eq.(20) \hspace{.1cm} \right ]\end{equation}

\begin{equation}
\hspace{-1cm} c = c_0 \left(1 +\frac{gh}{c^2}\right)\end{equation}}

Thus, if $ c_0$ be the speed of light at the place with zero gravitational potential, then Eq.(24) gives the speed of light $c$ at a place with gravitational potential $gh$. It is clear that $c > c_0$  only if  $ gh > 0$. So, when an observer measures the speed of light at his own position, the constancy of its speed holds. \\

Different rates of time flow also cause a light beam to take longer to cross a more gravitationally-dense region than one in which the background gravitational field intensity is weaker. This fact has been experimentally established and is called Shapiro effect [20].

  \section{Conclusion and discussion}
  
  \hspace{.5cm}Thus based on behavior of an atom and its logical consequences in an inertial moving frame and in gravitational field, we have explained both velocity time dilation and gravitational time dilation respectively. Although the velocity time dilation is caused by the virtue of inertial motion and gravitational time dilation by the virtue of gravitation but as we analyzed, both have essentially similar principles underlying them and similar procedures (like above) can be used to explain them. This make us infer that both kinds of time dilations are only different manifestations of a single phenomenon. \\
  
  Although we can deduce time dilation in an inertial moving frame from the constancy of light speed as we do traditionally, however we must note that light speed constancy is not the cause of time dilation or in other words time dilation is not the consequence of light speed constancy. However vice versa of this can be true. The reason is that time dilation does not occur only in a photon clock but can occur in all other kinds of clocks as well, including complex ones such as decaying radioactive particles or even biological systems (where light is not involved). Thus we can say that constancy of light speed is not a prerequisite for time dilation. Infact, we can consider a moving photon clock as a particular case in which time dilation causes constancy of light speed as explained in this paper. Einstein's question [21] in this connection is note worth
\begin{quote}
``{\itshape Can we not assume such changes in the rhythm of the moving clock and in the length of the moving rod that the constancy of the velocity of light will follow directly from these assumptions?} " 
\end{quote}

Now if we place the photon clock in a lower gravitational potential, it will tick slower due to gravitational time dilation, but the path length of light between the mirrors does not increase. Therefore light will take more time to travel the same distance between the mirrors and thus its speed will appear to decrease for the observer who is in higher gravitational potential. Thus in this case it is time dilation which affects speed of light and not speed of light which causes time dilation. Time dilation is more general phenomenon of nature than light speed constancy.\\

\vspace{1.5cm}Lorentz transformation (LT), which were derived by Joseph Larmor [22] in 1897, and Lorentz (1899, 1904) [23], directly predicted time dilation. Infact time dilation by the Lorentz factor was correctly predicted by Joseph Larmor (1897), at least for electrons orbiting a nucleus, when he wrote [24] 
\begin{quote}
	``{\itshape...individual electrons describe corresponding parts of their orbits in times
	 shorter for the [rest] system in the ratio $\sqrt{1-\frac{v^2}{c^2}}$} "
\end{quote}

  However, when Einstein published his paper [8]`On the Electrodynamics of Moving Bodies ' in 1905, constancy of light speed had  gained experimental support where as time dilation was not yet verified experimentally. But now we have strong experimental evidence for both velocity time dilation [25][26][27] and gravitational time dilation [28][29].\\
  
  Since 1905, lot of theoretical work has been done concerning special relativity. LT which show time dilation by Lorentz factor can very well be derived without constancy of light speed by pure algebra. After Minkowski [30] addressed the space-time geometrical nature of the LT, it was soon realized, that the area of applicability of the LT extends beyond the classical electrodynamics. Constancy of light speed is not necessary for its derivation [31][32].     Numerous derivations of the LT are published that do not require the constancy of light speed as a postulate [33]. Infact, the over emphasized role of speed of light in the foundations of the special relativity has been even subjected to criticism [34].\\
  
  To consider time dilation as a more fundamental phenomenon of nature and constancy of light speed as one of its consequence in a particular case will have advantages. It will allow us to view, velocity time dilation and gravitational time dilation, together as one in essence. This approach will make special relativity intuitionally more acceptable. Finally it will not only provide us with an answer to the question that why is speed of light constant in an inertial frame but will also endow us with a clue for its much sought mechanism !\\\\

  \vspace{2cm}{\large \bfseries Acknowledgement }\\\\I am indebted to my mentor Prof.SMA Hashmy for his guidance. I thank J.H.Field for his useful comments. I also thank my friend Mir Faizal for useful discussions. I am grateful to David J. Griffiths for his encouragement. I am highly gratified for the colossal love and support of my brother Dr.Muhammed Banday and my mother Mrs.Mehmooda.  At last, I offer my salutation to one of the greatest physicists of all time - Ibn al-Haytham, who has ever been and continues to be my inspiration.


\begin{thebibliography}{28}
\bibitem{}  NIST and BIPM practice is to use $ c_0 $ for the speed of light in vacuum in accord with international standard ISO 31-5.However throughout this paper $c$ will be used exclusively to denote the speed of light in free space, unless otherwise stated.
\bibitem{} Length contraction implies that when a body moves with a velocity $v$ relative to the observer its measured length is contracted in the direction of its motion by the factor $ \sqrt{1-\frac{v^2}{c^2}}$, whereas its dimensions perpendicular to the direction of motion are unaffected. Thus the distance between mirrors will decrease from $L$ to $L \sqrt{1-\frac{v^2}{c^2}}$ as the angle between clock mirrors and direction of its motion will vary from $0^\circ$ to $90^\circ$.\\\\
\end{thebibliography}
\end{document}